\documentclass[apj]{emulateapj}

\shorttitle{M87 Globular Clusters}
\shortauthors{Harris }

\begin{document}

\title{The Globular Cluster System in M87: A Wide-Field Study with 
CFHT/Megacam\footnote{This research used the facilities of the Canadian Astronomy Data Centre
operated by the National Research Council of Canada with the support of the Canadian
Space Agency.  Based on observations obtained at the Canada-France-Hawaii Telescope 
(CFHT) which is operated by the National Research Council of Canada, the Institut National 
des Sciences de l'Univers of the Centre National de la Recherche 
Scientifique of France,  and the University of Hawaii.}
}

\author{William E. Harris}
\affil{Department of Physics \& Astronomy, McMaster University,
  Hamilton ON L8S 4M1}
\email{harris@physics.mcmaster.ca}

\clearpage

\begin{abstract}
CFHT Megacam data in $(g',r',i')$ are used to obtain 
deep, wide-field photometry of the globular cluster system (GCS) around  M87.  
A total of 6200 GCs brighter than $i'=23.0$ (roughly equivalent to $M_I = -8.5$)
are included in the study, essentially containing almost the entire bright
half of the total GC population in the galaxy.
The classic bimodal metal-poor and metal-rich sequences of GCs show
up clearly; while the spatial distribution of the GCs can be traced detectably outward to
$R_{gc} \simeq 100$ kpc and perhaps further, the blue, metal-poor subpopulation is 
very much more spatially extended than the red subpopulation.  
Both the red and blue GC subsystems have radial metallicity gradients,
where mean heavy-element abundance scales  with projected galactocentric
distance as $Z \sim R^{-0.12}$ (blue) and $R^{-0.17}$ (red).
The blue sequence exhibits a strongly significant
mass/metallicity relation (MMR) in which the mean metallicity gradually
increases with cluster luminosity as $Z \sim L^{0.25 \pm 0.05}$ for the
luminosity range $M_I \lesssim -10$ and the
assumption of a constant $M/L$.  However, this relation is also clearly
nonlinear:  fainter than this level, the sequence is more nearly vertical.
This mass/metallicity trend can be understood as the result of self-enrichment
within the most massive metal-poor GCs during their formation.
The red sequence formally exhibits a negatively sloped MMR, but the 
numerical solutions and tests show that this red-GC slope is not very significant.
In giant elliptical galaxies, the red GCs are likely to represent a broad composite 
population formed during several major starbursts. If so, the red sequence
might display a \emph{population-based} MMR that could in principle be either 
positive or negative.

\end{abstract}


\section{Introduction}

By far the largest collection of globular clusters (GCs) within any one galaxy
in the local universe is in M87, the cD giant in the Virgo cluster.
Its total GC population is $N \simeq 14000$ \citep{mcl94,tam06}, 
roughly a third of the GCS population in all the Virgo galaxies combined \citep[see the Virgo Cluster
Survey results of][]{pen06}.  In addition, because both the
foreground reddening and field contamination are low, the M87 GCS     
can be traced outward to extremely large projected galactocentric distance,
100 kpc or more \citep{tam06}.
This combination of characteristics makes M87 a unique testbed for
systematics of GC systems.

A particular topic of recent interest is the correlation between
GC metallicity and luminosity (or mass), a trend which has been found to
show up on the blue, metal-poor sequence 
\citep[][and references cited there]{har06,str06,mie06,spi06,for07,weh08,bas08,har09}.
In brief, the mean GC metallicity along the blue sequence is seen
to increase steadily, though perhaps not linearly, with increasing
cluster mass.  So far, the most likely physical explanation 
that has been advanced for the origin of this effect 
is some form of self-enrichment during
the protocluster epoch \citep{har06,mie06,str08,bai09}. 
The essential idea of this interpretation is that any self-enrichment should be more
efficient within the more massive, deeper potential wells of the
larger clusters.

M87 has turned into something of a flashpoint for discussion of
this mass/metallicity relation (MMR).  \citet{str06} first claimed
that the mean GC color along the entire blue sequence increased 
roughly linearly with absolute magnitude, which would correspond
to a simple power-law scaling of heavy-element abundance with
cluster mass, $Z \sim M^p$, for a constant mass-to-light ratio.  
Their measurements consisted of simple
fixed-aperture photometry of $\sim1000$ GCs around M87 from the
single-orbit exposures in $(g',z')$ of the galaxy that
were taken as part of the Virgo Cluster Survey
\citep{cot04}.  \citet{mie06} used the same material in
the full Virgo Cluster Survey analysis in which the
cluster photometry was done with PSF-convolved King profiles.
They found much the same trend though it was less clear that 
the MMR was as straightforward as a linear fit in the color-magnitude plane.
\citet{for07} used ground-based $C,T_1$ photometry with standard
PSF-fitting photometry, and found a roughly linear correlation
of $(C-T_1)$ color versus magnitude along the blue sequence.
\cite{wat09}, on other hand, find no net MMR along the blue
sequence (that is, they find a vertical sequence of magnitude
versus color) by using size-corrected aperture photometry of
about 2000 clusters from deeper HST/ACS images of M87 in $(V,I)$.
Most recently, \citet{pen09} have analyzed the same ultra-deep
HST/ACS data with the PSF-convolved King profile technique
described in \citet{jor09} to find that a significant color/magnitude
slope exists for the upper half of the GC magnitude range
(roughly, for the clusters brighter than the turnover point
of their luminosity function).
Thus five different measurement procedures from four 
partially different datasets have yielded results that 
are largely in mutual agreement, but also show points
of disagreement that are still of some importance and need to be settled.

For comparison,
the largest GC database used specifically to analyze the MMR is
the sample of $\sim 10,000$ clusters in six giant Brightest Cluster
Galaxies \citep{har09}.  There, the photometry of the individual GCs
is obtained from PSF-convolved King-model profiles and is taken in
the metallicity-sensitive $(B-I)$ index and with deep 
exposures.\footnote{In this and most previous MMR work, the
fundamental but well justified 
assumption is made that the spread of broadband optical
color indices directly represents differences in metallicity,
with [Fe/H] increasing monotonically as the color index gets redder.
The effects of other parameters on their broadband colors
(such as cluster age or $\alpha-$abundance)
generate far more subtle changes in color.
Convincing measurements of age or $\alpha$ differences between old GCs 
traditionally require
careful analysis of high$-S/N$ integrated spectra and are 
challenging to uncover even then; see, for example, 
\citet{puz05,pen06,str06,pes08} among others, as well as comparisons
of contemporary simple stellar
population (SSP) models \citep[e.g][]{pes08}.}
For clusters brighter than the GCLF turnover 
the data have high S/N, and the total sample is so
large that the GCLF can be traced upward to the most
luminous known GCs at $L \gtrsim 10^7 L_{\odot}$.
Extensive evidence is presented that the MMR has a
nonlinear form:  the blue GC sequence 
is nearly vertical (i.e., no trend of GC color versus luminosity) 
for $L \lesssim 5 \times 10^5 L_{\odot}$
and then curves toward redder color at higher luminosity.

The quantitative self-enrichment model of \citet{bai09} 
reproduces this nonlinear MMR form fairly accurately. The key physical point
in the model is that protoclusters less massive than about
$10^6 M_{\odot}$ do not have deep enough potential wells to
hold back much of the first round of SNII-enriched gas, so for these
lower-mass GCs there should be no MMR on \emph{either} the blue or
red sequence.  Conversely, for the higher-mass GCs with their deep
potential wells, self-enrichment is almost automatically required.
A unique prediction of this particular model is that \emph{both}
the red and blue GC sequences should exhibit a MMR at the top end,
though it should be less noticeable for the red sequence because
it starts from a much higher level of pre-enrichment.

Settling the issue \emph{on observational grounds} of 
the detailed form of the MMR (and indeed whether or not it exists
in all galaxies) is of some urgency because 
confusion over the form of the correlation has already arisen
in the literature.  Yet
it is likely to bear quite directly on 
understanding the formation of massive GCs and their early
star formation history \citep{har06,mie06,str08,bai09}. 

In this paper I discuss new photometry of the M87 GCs obtained 
with CFHT/Megacam. The data comprise a large enough GC sample ($N \simeq 10^4$)
to analyze the MMR to deep limits in addition to correlation
with galactocentric distance.  The sample size 
provides the obvious advantage that features such as the MMR,
the spatial distribution, or the GCLF can
be studied within a \emph{single} galaxy without having to put
together composite samples.  The results discussed in the
following sections show well defined blue and red
GC sequences extending to high mass, and exhibit a nonlinear MMR
quite similar to the ones in the six-galaxy BCG sample.  Though M87 alone
can hardly settle the question of the universality of the MMR, 
these results reinforce
the need to investigate this intriguing effect further.
In the present paper, I concentrate on the various features
of the color and metallicity distributions.

In this discussion I adopt a distance modulus $(m-M)_0 = 31.0$
for M87 and the core of the Virgo cluster \citep{jer04,cal06,wil07}
and a foreground absorption and reddening $A_I = 0.04$, 
$E_{V-I}=0.03$.  In the SDSS filter
system the corresponding values relevant here
are $A_i = 0.04$, $E_{g-i} = 0.04$, $E_{g-r} = 0.02$.

\section{The Data Sample}

The 36 CCDs making up the CFHT Megacam array provide a $1^o$ square
field of view and the camera is normally used with the $u'g'r'i'z'$
filter set (in what follows I will drop the accents and refer to
the magnitudes simply as $ugriz$).  The image scale is $0.187''$/px
and the entire field makes up a 20K $\times$ 20K array.
The data used in this paper consist of a single set of 
homogeneous exposures centered on M87 taken in $gri$ during 
2004 March and April (Harris, PI), as listed in Table 1.  
While this material will eventually be surpassed by the much
more extensive CFHT/Megacam Next-Generation
Virgo Cluster Survey by Ferrarese et al. (see www.astrosci.ca/NGVS),
much can already be extracted from the earlier set of M87-centered
exposures.

Individual exposures in each filter were taken
with a ``small dither'' pattern sufficient to splice over the small gaps
between the individual CCDs in the array, but the much larger gaps between
the top two ranks of CCDs and the bottom two ranks are still present
(standard large-dither patterns are available for the camera that would
cross these gaps, but these make
the astrometric re-registration more difficult).  These remaining gaps show
up as two $\simeq 350-$pixel-wide blank E/W stripes in the array,
making up about 3.5\% of its total area on the sky.
The hour-long total exposure times in each filter were chosen to reach closely similar
limiting absolute magnitudes for the globular clusters.
The combined images in each filter are the ones 
produced through the Megapipe facility of CADC,
and were downloaded from the CADC Archive ready for photometry.  

The GCs around M87 are seen against a population
of field objects including foreground Milky Way stars and background
galaxies.  Because the Virgo field is at high latitude, the
contamination from stars is minimal and the main contaminants are
small, faint galaxies.  However, 
much of the field galaxy population can be removed by morphology 
because the GCs themselves appear starlike
on these ground-based images.
At the 16-Mpc distance of Virgo, an \emph{average} GC half-light
diameter of 5 pc \citep{har96} is equivalent to $0.06''$, which 
in turn corresponds to a FWHM of $0.04''$ for an average King-model
concentration ${\rm log} (r_t/r_c) = 1.5$ \citep{lar99}.  This scale
size is thus 20 times smaller than the
$0.8''$ FWHM seeing on the combined Megacam frames in this
dataset, making any such objects 
indistinguishable from stars.  Extensive quantitative tests of
PSF-convolved globular cluster profiles for distant galaxies such as
these show explicitly that the intrinsic profile width falls below
the threshold of measurement if its FWHM is less than 10 percent of
the PSF width \citep[e.g.][]{lar99,har09}.  In practice, what this 
means for the present data is that 
the GCs can be accurately measured either through conventional fixed-aperture
or PSF-fitting photometry.  Because the crowding levels are minimal,
aperture photometry is adopted here (see below),
with an aperture radius set at its optimal value similar to the 
PSF FWHM.  Use of aperture photometry also alleviates concerns over
minor changes in the PSF shape across the field.

As another basis of comparison, it is worth noting that imaging of the
M87 clusters with \emph{ground-based} image quality of $0.8''$ is almost
exactly analogous to imaging clusters in the Coma galaxies at $d=100$ Mpc
with \emph{HST} image quality of $0.1''$.  Explicit tests of PSF-convolved
GC profiles for the Coma galaxies \citep{har09b} with deep HST data verify that
the intrinsic sizes of the GCs are too small to affect their integrated 
magnitudes from either PSF-fitting or fixed-aperture photometry at
levels larger than $\gtrsim 0.02$ mag.  Even further,
when these second-order magnitude corrections are differenced between 
two filters ($g-i$ or $g-r$ in this case), the effects of aperture size corrections 
on \emph{color} are even smaller third-order ones that can be completely neglected.
Similar comments apply to the $(C-T_1)$  ground-based photometry of M87 and NGC 1399
by \citet{for07} in which the seeing was $1.0''-1.6''$.

\begin{figure}
\plotone{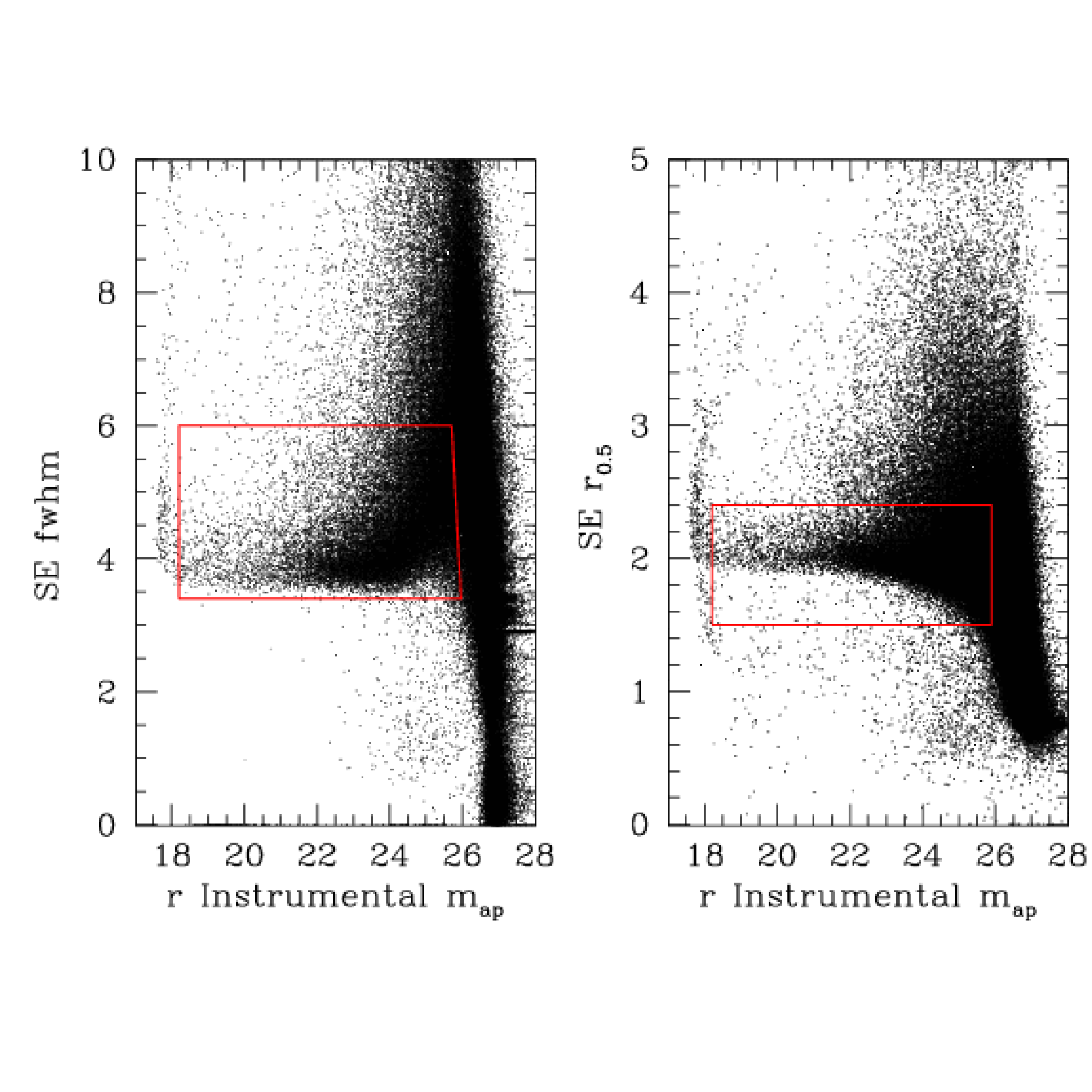}
\caption{Use of SExtractor parameters for rejection of nonstellar objects.
The measured FWHM and effective radius of each detected object on the
M87 field are plotted against aperture magnitude.  Objects lying outside
either of the marked boxes are rejected from the detection lists (see text).
The example shown here is from the $r-$band frame; note that the $r$ values 
plotted here are only the raw instrumental magnitudes and should not
be compared with the final color-magnitude diagrams below.
}
\label{segraphs}
\end{figure}

The first step in the photometry was to find candidate objects on all
three master images and to eliminate the majority 
of nonstellar objects.  The SExtractor code \citep{ber96} was 
run to do this, and plots were generated of the SE effective
radius and FWHM versus aperture magnitude, as shown in 
Figure \ref{segraphs}.  Objects falling outside the generous ``stellar'' boxes
marked in the figures are either very bright, saturated stars;
obviously nonstellar extended galaxies; or extremely faint objects,
all of which could be eliminated from the detection lists.

Next, the SE-culled finding lists defined separately in $g,r,i$ were
compared, and any objects found on \emph{all three} filters and
coordinate-matched to within $\pm 1.5$ px were kept for the final
photometric measurement run.  These matched objects were measured through
\emph{daophot} with an aperture radius of 3.5 px ($0.65''$, equivalent to a linear
size of 50 pc at the distance of M87).  Crowding in this
high-latitude field is negligible, and comparisons of the fixed-aperture
magnitudes with PSF-fitted magnitudes showed no differences in the resulting
color-magnitude diagrams.

\section{Astrometric and Photometric Calibration}

\begin{figure}
\plotone{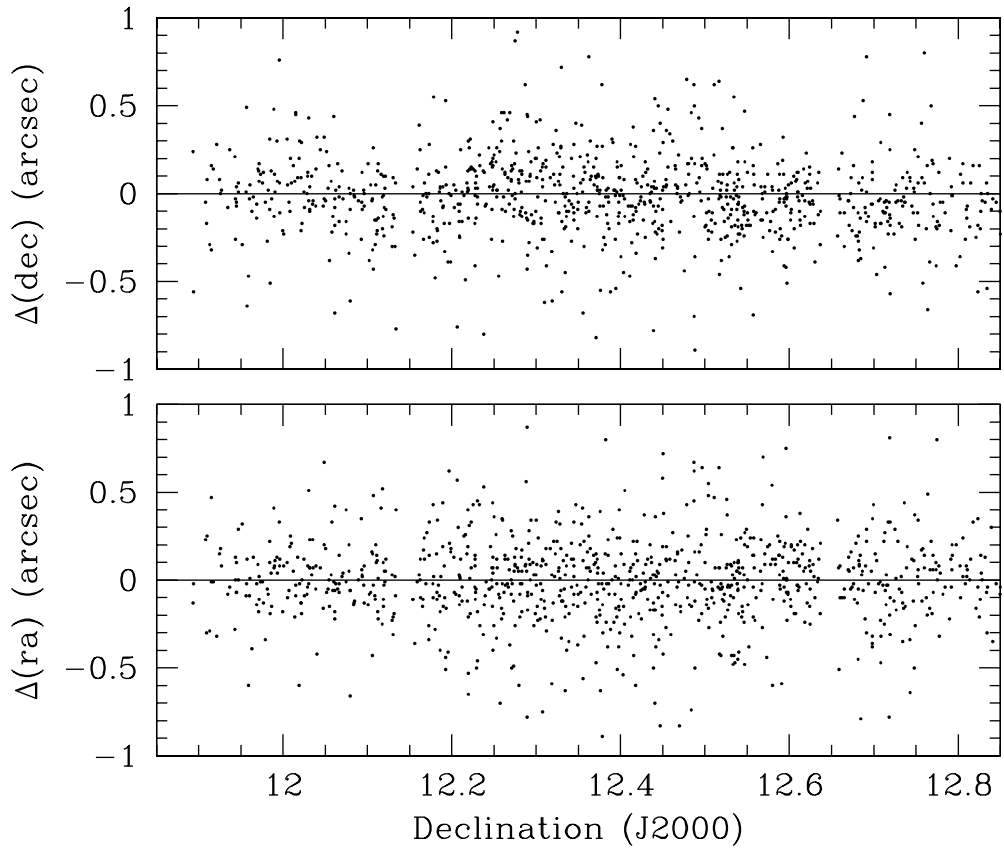}
\caption{Residuals in the measured positions for the 932 stars in the
USNO-B1 catalog that were found to lie within the M87 Megacam field,
plotted as a function of declination in degrees.  The rms scatter around
zero is $\pm 0.23''$ in both right ascension and declination.
}
\label{astrom_resid}
\end{figure}

The Megacam stacked images in the CADC Archive have internally
homogeneous astrometry that is already close to the most commonly
used systems (such as SDSS, 2MASS, USNO), but to ensure that the data were
on a specific astrometric system as closely as possible, the
USNO-B1 system was selected.  The USNO-B1 catalog
was searched for all objects within the Megacam field,
and the positions of these objects were matched up with the $(\alpha, \delta)$
values returned directly from the Megacam image headers.   This matchup
yielded 932 stars in common to within $1''$.  Small zero-point
offsets equal to $\Delta\alpha = (0.053\pm 0.008)''$, 
$\Delta\delta = (0.254\pm 0.008)''$ in the sense (USNO-CFHT) were found, but no
significant higher-order terms.  Figure \ref{astrom_resid} shows
the residuals \emph{after} the native Megacam astrometric coordinates
were shifted onto the USNO-B1 system.  The object-to-object rms scatter
is $\pm 0.23''$ across the entire field, with no systematic deviations
larger than $\pm 0.05''$.

The Megacam $ugriz$ magnitudes are known to differ slightly from the Sloan 
Digital Sky Survey scales (see the Megacam webpages for
updated calibration information).
However, for the purposes of more
direct comparison with other GCS data in the literature, it is useful
to put the present M87 data onto the SDSS scale.  The SDSS DR5 catalog was searched
for objects in the M87 field and compared with the Megacam object lists, 
with the result that matches
were found for about 1000 stars brighter than $i=21.0$.
These were used directly to set the photometric zeropoints for $gri$ in
the M87 field.  The mean differences $\Delta m = m(CFHT)-m(SDSS)$ were found to be
$\Delta(g) = 0.254 \pm0.004$, $\Delta(r) = 0.407 \pm0.008$, and 
$\Delta(i) = 0.412 \pm0.006$, where the CFHT magnitudes are the ones defined
simply using the zeropoints in the image headers.

A secondary advantage of tying these data closely to a widely used standard
system is that the intrinsic colors of globular clusters in the 
$(g-r)$ and $(g-i)$ indices have not previously been well established.  The Virgo Cluster
Survey data are in $(g-z)$ through the HST/ACS filters
\citep{pen06,mie06}.  Previously published
studies of GCSs for various galaxies in $(g-i)$ include \citet{for04}, \citet{weh08}, and
\citet{coc09} all taken with Gemini/GMOS, 
but in all of these the absolute photometric calibrations are preliminary
and not likely to be accurate to better than $\pm0.1$ mag (see these papers
for detailed discussion).  M87 offers an excellent opportunity to set up
a fiducial color scale for GCs, not just because a direct matchup to
the SDSS catalog is possible, but also because its foreground reddening is almost
negligible and the statistical size of the GC population allows the
mean colors of the red and blue sequences to be defined internally
to $\pm 0.02$ mag or better (see below).

\section{Definition of the Globular Cluster System}

\begin{figure}
\plotone{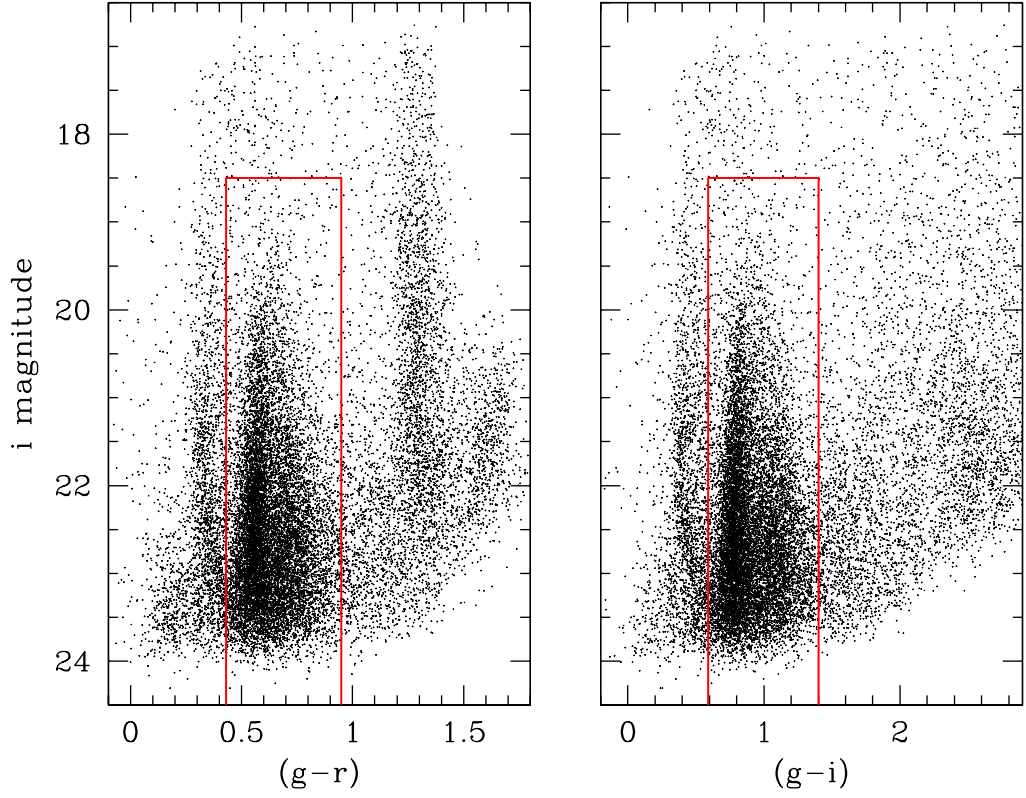}
\caption{Final color-magnitude diagrams for the starlike and near-starlike
objects detected in all three frames $(g,r,i)$. Objects lying outside either
of the marked boxes are rejected as field contamination (mostly foreground
Milky Way halo or disk stars).
}
\label{cmd_raw}
\end{figure}

In Figure \ref{cmd_raw} the final data for 20189 starlike 
or near-starlike objects are shown, plotted 
in $(i,g-r)$ and $(i,g-i)$.  The detection rate for objects within
the color range $(g-r) < 1.0$ is highly complete for $i < 23.5$; in the
following analysis, the main conclusions rely even more conservatively
on the interval $i < 23.0$ and are unaffected by any incompleteness.

Several clearly distinct components show up.
The GCs that we are after lie on the two nearly vertical, narrow 
sequences at $(g-i) \simeq 0.8, 1.05$ while the populations bracketing
these on the blue and red sides
are the expected loci of foreground Milky Way halo and disk stars.
Any objects falling in the regions
$$(g-r) < 0.43, \, (g-r) > 0.95$$
$$(g-i) < 0.59, \, (g-i) > 1.40$$
$$ i < 18.5 $$
(the boundaries of which are 
marked by the boxes in Fig.~\ref{cmd_raw}) are thus rejected as field contamination.
This final culling leaves 11946 objects that make up our final ``best'' GC candidate list.

\begin{figure}
\plotone{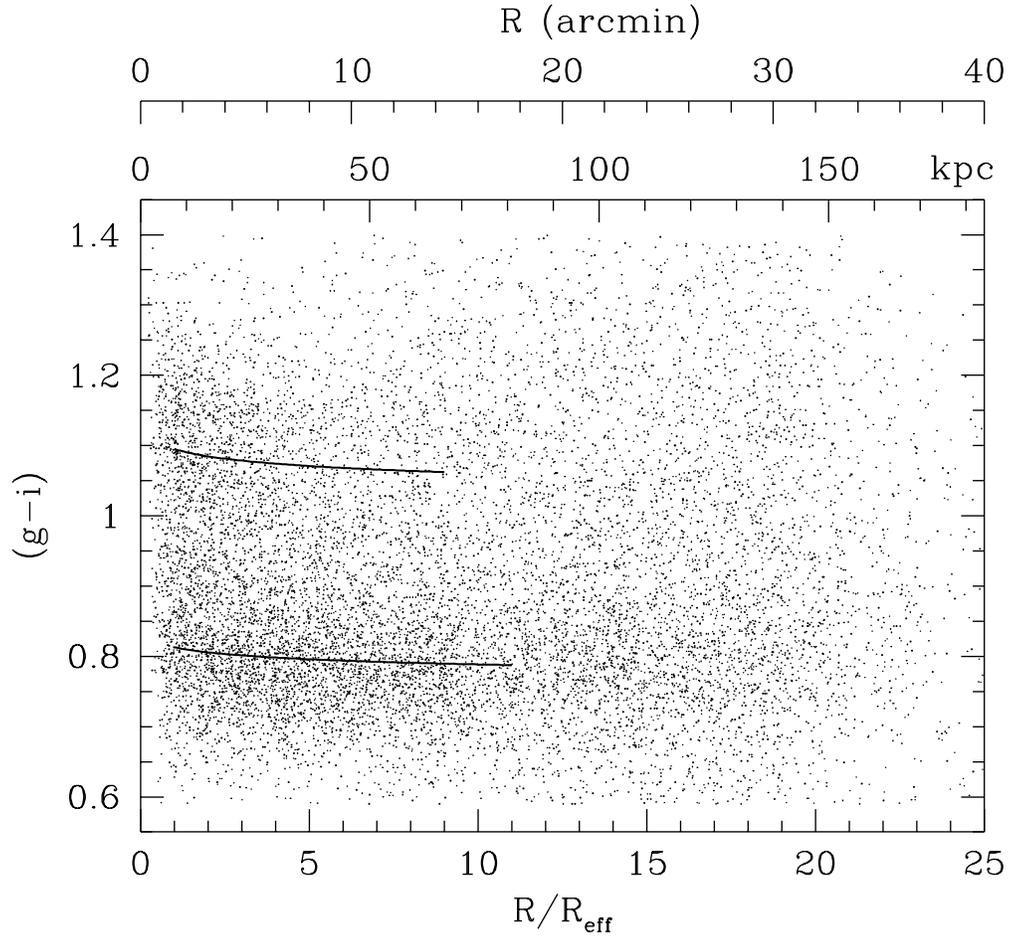}
\caption{Measured $(g-i)$ color index versus projected galactocentric
distance for the ``starlike'' objects brighter than $i=23.0$.  This sample
is dominated by globular clusters (see text).   Radii are plotted in
units of the M87 spheroid effective radius (bottom axis) or in
kiloparsecs or arcminutes (top axes).  The calculated radial 
metallicity gradients for the blue and red GC sequences are shown
by the solid lines.
}
\label{radcolor}
\end{figure}

The distribution of the GC candidate colors by spatial location is shown in
Figure \ref{radcolor}.  Here, the projected distance $R$ from the center
of M87 is expressed in units of $R_{eff} = 1.58'$, the effective radius of the
galaxy's spheroid light. 
A well known though still striking feature
of the M87 GC system is its huge radial extent:  the
clusters extend detectably beyond even $R \sim 150$ kpc or 20 $R_{eff}$
(see next section).
The first hints of its huge GCS halo were already evident in the wide-field
starcount work of \citet{har76} and more clearly outlined 
with the CCD-based photometry of \citet{mcl94} and \citet{tam06}.
For comparison, however, the $V-$magnitude data of 
\citet{mcl94} could be used to trace the GCS out only to the
radial limits of their single-CCD field at $R(max) \simeq 8' \simeq 40$ kpc.
\citet{tam06} used a combination of four Subaru Suprime-Cam pointings 
in three colors ($BVI$) to
make up a surveyed region of $136' \times 27'$, about the same total area as
the single Megacam field but oriented along the E/W direction.  M87 itself
is in the SW corner of their westernmost SCam pointing, while
another Virgo giant NGC 4552 is in their third field pointing.  

\begin{figure}
\plotone{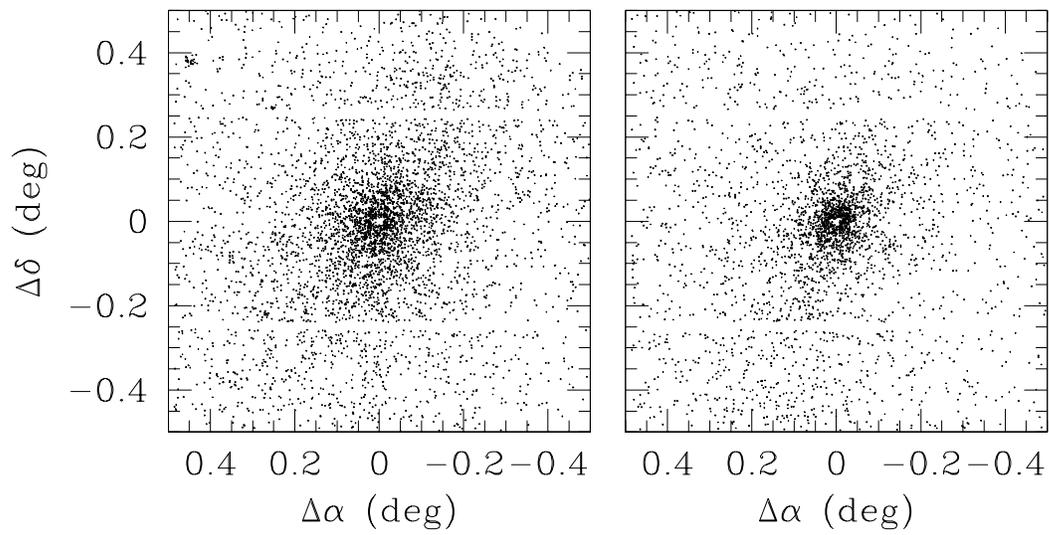}
\caption{Spatial distributions of GC candidate objects brighter than
$i=23.0$ as in the previous figure, subdivided into the blue sequence
(left panel) and red sequence (right panel).  The much larger spatial
extent of the blue GC population is evident.  Note also the two
vacant horizontal stripes above and below center that represent the
large gaps between the ranks of CCDs in the camera mosaic.
}
\label{xypair}
\end{figure}

\begin{figure}
\plotone{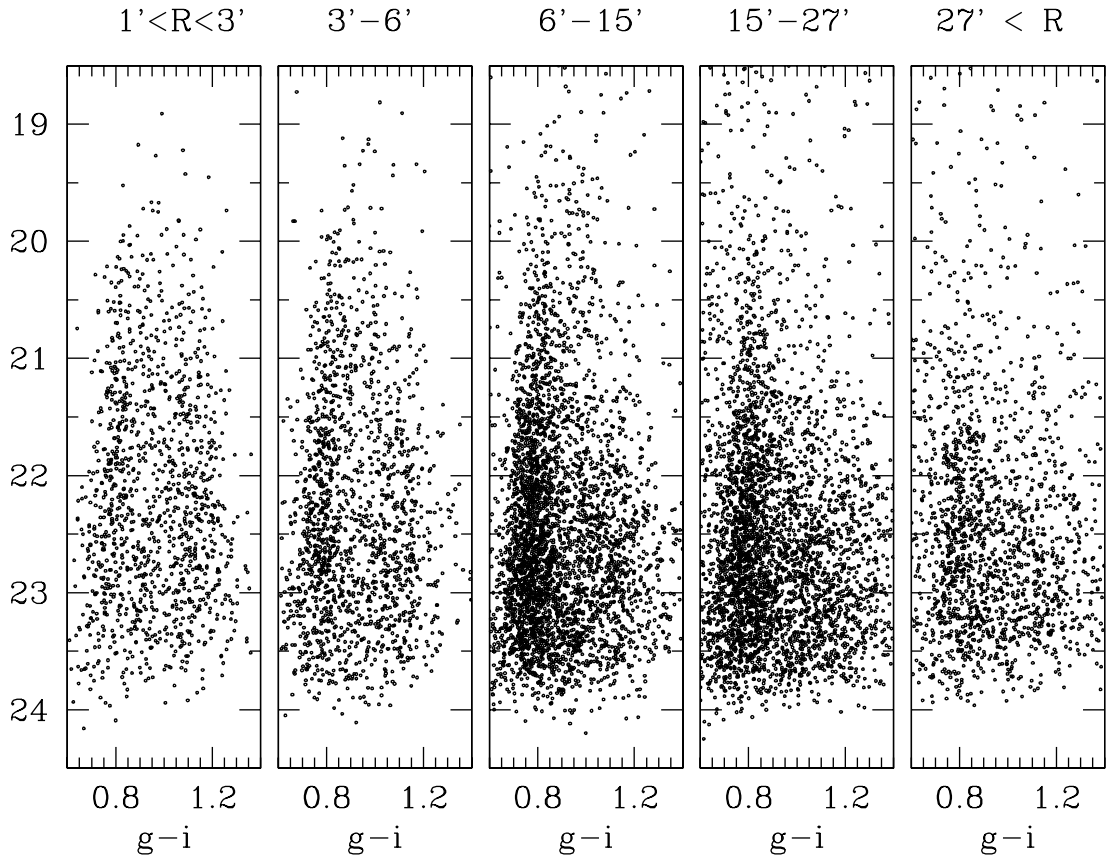}
\caption{Color-magnitude data in $(i,g-i)$ for the starlike objects, separated
by radial zone in projected galactocentric distance $R$.  Five zones are shown.  
}
\label{cmd_zone}
\end{figure}

An equally striking feature of the spatial distribution 
from Fig.~\ref{radcolor} is the difference
between the blue and red GC subpopulations.  The red sequence virtually
disappears into the general spread of field contamination for 
$R \gtrsim 6-8 R_{eff}$ ($\sim 12'$ or 50 kpc) while the blue 
sequence continues detectably
outward much further.  Two other illustrations of this are shown in
Figures \ref{xypair} and \ref{cmd_zone}.   
As a first approximation we define the blue GCs
as those in the color interval $0.59 \leq (g-i) \leq 0.93$,
and the red GCs as those in $0.93 < (g-i) \leq 1.40$.
In Fig.~\ref{xypair}, the spatial positions of each type are shown.
\citet{tam06} and \citet{for07} find exactly the same phenomenon 
and these papers provide a
more detailed discussion of the spatial distributions of each component.
In Fig.~\ref{cmd_zone}, the CMDs in $(i,g-i)$ are displayed for five
radial zones.  Whereas the blue and red sequences are almost equally
populated for the innermost zone $R < 3'$, the red sequence dies away 
more rapidly outward, leaving the blue sequence to dominate
at large radii.

\section{Radial Structure}

The M87 GCS is so populous and extended 
that the level of background contamination
(that is, the field contamination still present in
the list of starlike objects, assumed constant across the field)
is hard to determine precisely.  The contamination
is completely unimportant at
small radii, but at the largest radii of the survey it determines
the effective limit beyond which the GCS can be traced.
We have no truly remote ``control field'' adjacent to M87
for measuring $\sigma_b$, and ultimately a definitive measurement
of the background field level and the outermost radial profile
of the M87 GCS will have to await the still wider-field work
from the Next-Generation Virgo Survey.  For the present,
the outer parts of the single Megacam field itself must be used to gauge
$\sigma_b$.  A reasonable estimate can be made by noting that for
$R \gtrsim 12 R_{eff}$, the areal density of \emph{red} objects
($i \leq 23.0$ and $0.93 < (g-i) < 1.4$) is nearly uniform at
$\sigma = 0.36$ arcmin$^{-2}$,
accounting for the edges of the frame and the two blank stripes in
the CCD array.  The color range enclosing
the blue GC sequence is narrower ($0.59 \leq (g-i) \leq 0.93$),
so if the color distribution of the contaminants is roughly
uniform over this range, the level of the blue background will be
proportionately smaller, $\sigma_{bkdg}(blue) \simeq 0.26$ arcmin$^{-2}$,
for the same magnitude range $i < 23.0$.
(See also Harris 2009 for more extensive tests and discussion of the
background color distribution in similar high-latitude fields.)
Thus for the present purposes I adopt a 
total $\sigma_{bkgd} \simeq 0.6$ arcmin$^{-2}$,
with a true external uncertainty 
likely to be near $\pm 0.1$ arcmin$^{-2}$.

In the final list of GC candidates defined above, there are 8313
objects brighter than $i=23.0$.  Subtracting off the background,
this leaves a total GC population $N_{GC} \simeq 6200 \pm 350$
brighter than $M_I \simeq -8.5$, which is about 0.2 mag brighter than
the turnover (peak) point of the luminosity function.

\begin{figure}
\plotone{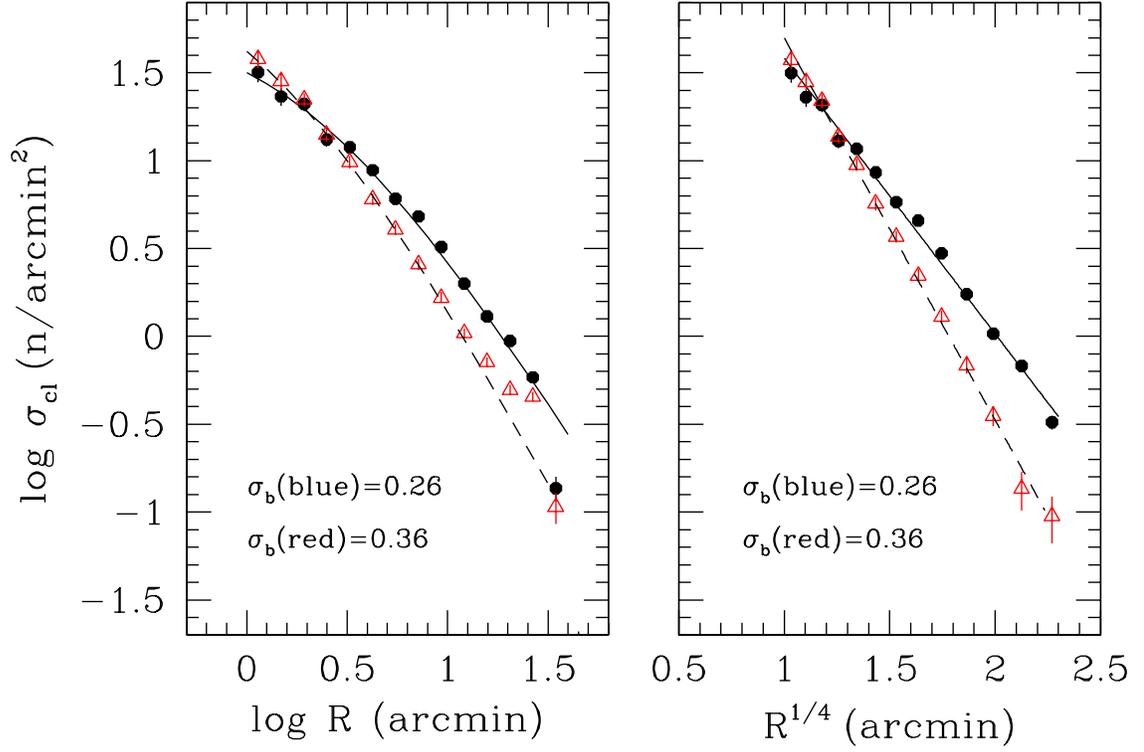}
\caption{Projected radial profiles for the blue and red GC subsystems,
using the $i< 23.0$ data from Figure \ref{radcolor}.  The number density
of GCs, $\sigma_{cl} = \sigma - \sigma_{bkgd}$, is plotted in power-law
form (left panel) and again in the de Vaucouleurs-law form (right panel).
The blue-sequence GCs are shown as solid dots and the red-sequence GCs
as open triangles.
In the left panel the Hubble profiles fitting each of the two components
as given in the text are superimposed.
}
\label{radprof}
\end{figure}

The actual radial profile of both the blue and red GC 
subsystems can be quantified
by tracing the projected surface density $\sigma_{cl}$ directly
as a function of radius.  The observed number density
of all the \emph{candidate} GCs is by definition $\sigma = \sigma_{cl} + \sigma_b$.
The resulting background-subtracted data for $i \leq 23.0$
are listed in Table \ref{radialdat} and 
shown in Figure \ref{radprof}.  The columns in the Table
give the inner and outer radii of each annulus, the mean
radius, the calculated total number of clusters in the annular
zone after background subtraction, and the density $\sigma_{cl}$.
The conventional
log/log form (log $\sigma_{cl}$ vs. log $R$) is shown in the left
panel, indicating that both the blue and red components fall off
progressively more steeply with increasing radius.
A standard Hubble profile,
which is a power law with a core radius $R_0$, 
\begin{equation}
\sigma_{cl}(R) \, = \, \sigma_0 \bigl(1 + {R \over R_0} \bigr)^{-a}
\label{hubble}
\end{equation}
provides a reasonable
fit to either component.  Fits of this model to the data (shown
in Fig.~\ref{radprof}) for the blue GCs give
($\sigma_0 = 66$ arcmin$^{-2}$, $R_0 = 2.0'$, $a=-1.8$);
while for the red GCs,
($\sigma_0 = 150$ arcmin$^{-2}$, $R_0 = 1.2'$, $a=-2.1$).
These are equivalent to simple power laws $\sigma \sim R^{-n}$
where the exponent $n$ steepens continuously with radius,
\begin{equation}
\label{hslope}
n(R) \, = \, a \, { (R/R_0) \over (1+{R \over R_0}) } \, .
\end{equation}

Both components also follow a deVaucouleurs-law form 
(log $\sigma_{cl}$ vs. $R^{1/4}$) fairly well over their entire
observed run.  In this form 
the best-fit equations describing each one are
\begin{eqnarray}
{\rm log} \sigma_{cl}(blue) \, = \, (3.154\pm0.067) \, - \, (1.760 \pm 0.045) 
(R/R_{eff})^{1/4} \, , \\
{\rm log} \sigma_{cl}(red) \, = \, (3.874\pm0.054) \, - \, (2.439 \pm 0.037) 
(R/R_{eff})^{1/4} \, .
\end{eqnarray}
where again the $\sigma$'s are in units of arcmin$^{-2}$ and the slopes
are normalized to the $R_{eff}$ of the halo light.
(These lines should, however, not be extrapolated inward where the
density distribution flattens off more strongly.)
For comparison, \cite{tam06} found slopes ($-1.47\pm0.03$)(blue) and
($-2.44\pm0.06$)(red).  As they discuss more extensively, the integrated
light of the M87 halo has a $R^{1/4}-$law slope of $-2.36\pm0.09$,
strongly suggesting that the more metal-rich GC population can be
directly linked to the similarly metal-rich population of the field
halo stars.  Much other discussion of similar
results from several other galaxies reinforcing this interpretation
and connecting with their evolutionary histories can 
be found in \citet[][]{lee98,lee08,har02,bas06,for07,shi09} among others,
and will not be repeated here.

The behavior of the deVaucouleurs-type profile fit itself, 
near its outer edge, provides a useful consistency
check of our adopted backgrounds $\sigma_b$(blue,red).  If we have
chosen the background nearly correctly, then the resulting $\sigma_{cl}$
profile can plausibly be claimed  
(as is the case for most other elliptical galaxies)
to follow a nearly straight line in the log $\sigma_{cl}$ 
vs.~$r^{1/4}$ plane all the way to the outer edge of the data.
An overestimate of $\sigma_b$ would cause the background-subtracted
profile to drop suddenly and unrealistically to zero at the edge,
whereas an underestimate would cause $\sigma_{cl}$
to flatten off artificially at large radius.  Neither one is
the case (Figure 7b), so the choice of background levels appears to be
self-consistent within their stated internal uncertainties.

The $\sigma_{cl}$ data indicate that the blue GCs exist in
detectable numbers out to at least $R=28'$ or $\simeq 130$ kpc.
A final check of the present results comes from comparison with
\citet{tam06}, whose SCam data extend along a line towards NGC 4552
out to $120'$ (more than 500 kpc) from M87.  
They find the blue-sequence GCs to extend detectably to $R=30'$
(see their Figures 5 and 7), and present a model fit to the
combined M87/N4552 counts that suggests the M87 GCs could even
extend to $100'$ and more.  Clearly, we have not yet traced the
full extent of this system.

\section{Metallicity Gradients}

A second-order feature of the radial color distributions (Fig.~\ref{radcolor})
is a progressive change in the mean color of both the red and blue GCs
with radius:  in other words, both components show a metallicity gradient.
This feature is clearest for $R \lesssim 8 R_{eff}$ or about 60 kpc; beyond 
that point, the red clusters are almost absent and the blue clusters show
no significant change in mean color.  For the region inward of $8 R_{eff}$,
power-law fits yield
\begin{eqnarray}
\langle g-i \rangle(blue) \, = \, 0.813 - 0.025 {\rm log} (R/R_{eff}) \, ,\\
\langle g-i \rangle(red) \, = \, 1.095 - 0.035 {\rm log} (R/R_{eff}) \, .
\end{eqnarray}
where both slopes are uncertain by 10 percent.
The single-stellar-population models of \citet{mar05}, which provide integrated color
indices for a variety of broadband colors including the SDSS filters,
show that $\Delta(g-i)/\Delta[Z/H] = 0.21 \pm 0.05$ over the metallicity
range [Z/H] $\lesssim -0.2$ that covers most GCs.  The color slopes then
convert into metallicity gradients $Z \sim R^{-\gamma}$ 
where $\gamma = 0.12\pm0.02$ (blue) and $0.17\pm0.03$ (red).

Comparable data for the GCSs in
several other large galaxies are now in the literature.
\citet{gei96} found $\gamma = 0.15 \pm 0.03$ (blue), 
$0.12 \pm 0.06$ (red) in the other Virgo supergiant NGC 4472.
\citet{for01} found $\gamma \sim 0.2$ for the blue GCs in two
Fornax ellipticals, NGC 1399 and 1427.  Most recently, \citet{har09}
derived a mean $\gamma = 0.10 \pm 0.02$ for both the blue and red
GCs in six giant Brightest Cluster Galaxies.  All these results
are similar, and begin to present a very consistent picture:
the mean GC metallicity in these large
galaxies depends on spatial location, becoming more enriched
the further inward it resides in the halo.

The widespread presence of metallicity gradients, albeit shallow,
indicates that \emph{both the major epochs of GC formation}, metal-poor
and metal-rich, had a significant component of dissipative collapse
of protogalactic gas and \emph{in situ} formation which led to
higher enrichment levels deeper in the potential wells of
these large galaxies.
Although gradients in the range $\gamma \sim 0.1 - 0.4$ for
the dominant \emph{metal-rich} field-star component have been
found in many ellipticals within their central bulges ($R \lesssim R_{eff}$),
the globular cluster data indicate that dissipative formation extended
much further outward into their halos, to $\sim 5 R_{eff}$ and beyond.
What is perhaps equally important for understanding galaxy formation
is that the \emph{metal-poor} component, which is all but undetectable
in the integrated field-star light, also has an inbuilt metallicity
gradient to the same degree.  This result makes it unlikely that
the metal-poor stars all formed in pregalactic dwarfs that only
later began to merge into bigger galaxies.  Instead, the first
major metal-poor round of star formation must have taken place 
while the initial stages of hierarchical merging were already
underway \citep[see][for further discussion and references]{har09}

\section{Bimodal Sequence Fitting}

We now look more closely at the blue and red GC sequences in the
color-magnitude diagrams.  The goal is to characterize each of the two sequences 
through their mean colors \emph{as a function of magnitude}
(the MMR) and their internal color (metallicity) dispersions. 
To define the MMR (or lack of one) we adopt the same basic approach as used
in several previous studies \citep[e.g.][among others]{har06,mie06,weh08,bas08,har09}:  
the CMD is divided into
narrow magnitude bins, and within each bin the color distribution of the
GCs is fitted by a bimodal Gaussian model.  Because each magnitude bin
is treated independently, no \emph{a priori} assumptions are made about the
particular form of the MMR, whether linear or otherwise.  

For the multimodal fitting, the
code RMIX is used here \citep[for extensive descriptions of RMIX applied to this
situation, see][]{weh08,har09}.  The default approach -- facilitated by the
sheer statistical size of the M87 GC population -- 
is to allow the code to solve freely for all five basic parameters in a Gaussian bimodal
distribution:  the mean colors of the two modes $\mu_1, \mu_2$ (blue, red), their
dispersions $\sigma_1, \sigma_2$ (blue, red), and the relative numbers
(proportions) $p_1/p_2$ of objects in each mode.  The MMR along either of
the sequences is then defined by any trend in the mean $\mu$ with magnitude.

The bimodal fits were done for objects within the zone $R < 6.5 R_{eff}$, i.e. within
the part of the M87 halo where both blue and red sequences are definitely present
and in roughly equal numbers (Fig.~\ref{radcolor}).  We restrict
the sample to objects in the color range $0.65 < (g-i) < 1.30$, and as a preliminary
step, the shallow metallicity gradient described in the previous section is
removed by normalizing the color indices to the values that all the clusters
would have if they were at $R = R_{eff}$.  That is, for the blue-sequence GCs (defined as those
objects with $(g-i) < 0.95$), we define $(g-i)_{norm} = (g-i) + 0.0245 {\rm log}(R/R_{eff})$,
and for the red-sequence GCs ($(g-i) > 0.95$), we define 
$(g-i)_{norm} = (g-i) + 0.0345 {\rm log}(R/R_{eff})$.

The results of these bimodal fits in both CMDs are summarized in
Table 2, 3, and 4.  Four separate runs were done with the following
conditions, which were chosen to allow additional tests of the reality of any 
deduced MMR along both of the sequences:
\begin{itemize}
\item{} The data were divided into 0.25-magnitude bins in $i$, and within
each bin all five free parameters ($\mu_1, \mu_2, \sigma_1, \sigma_2, p_1/p_2$)
defining the bimodal Gaussian fit were freely determined by the fit 
(i.e., unconstrained).  The exceptions to this
were at the bright end, where for $i < 20.5$ half-magnitude bins were used to
capture enough objects for a statistically significant fit.  In addition, 
preliminary runs showed that for the four brightest bins ($i < 21.0$) the 
fully unconstrained fits tended to be unstable, so for these, 
the intrinsic widths of the two sequences were constrained to be equal
to their average values on the lower bins,
$\sigma_1 \equiv 0.060$ and $\sigma_2 \equiv 0.10$.
The results for this run are summarized in Table 2.
The columns list the $i-$magnitude range defining each bin; the number of objects in the bin;
the best-fit blue-sequence parameters ($\mu_1, \sigma_1$) and their uncertainties;
the red-sequence parameters ($\mu_2, \sigma_2$) and their uncertainties;
and the best-fit proportion $p_1$ (the fraction of the total population belonging
to the blue sequence).  By definition $p_2 = 1-p_1$.
\item{} The same conditions as above were used, except that the \emph{red} sequence
was constrained to have a mean color $\mu_2 \equiv 1.070$, the
average obtained from the lower-luminosity bins in the first (unconstrained) run.
The color of the blue sequence was determined freely from the fit.
The results for this run are summarized in Table 3.
\item{} The same conditions were used, except that next the \emph{blue} sequence
was pinned at a mean color $\mu_1 \equiv 0.80$, the
average obtained from the lower-luminosity bins in the first (unconstrained) run.
The color of the red sequence was determined freely from the fit.
The results for this run are summarized in Table 4.
\item{} The mean colors of both sequences were pinned ($\mu_1 \equiv 0.80, \mu_2 \equiv 1.07$)
and only the $\sigma$'s and proportions were solved for.
\end{itemize}

\begin{figure}
\plotone{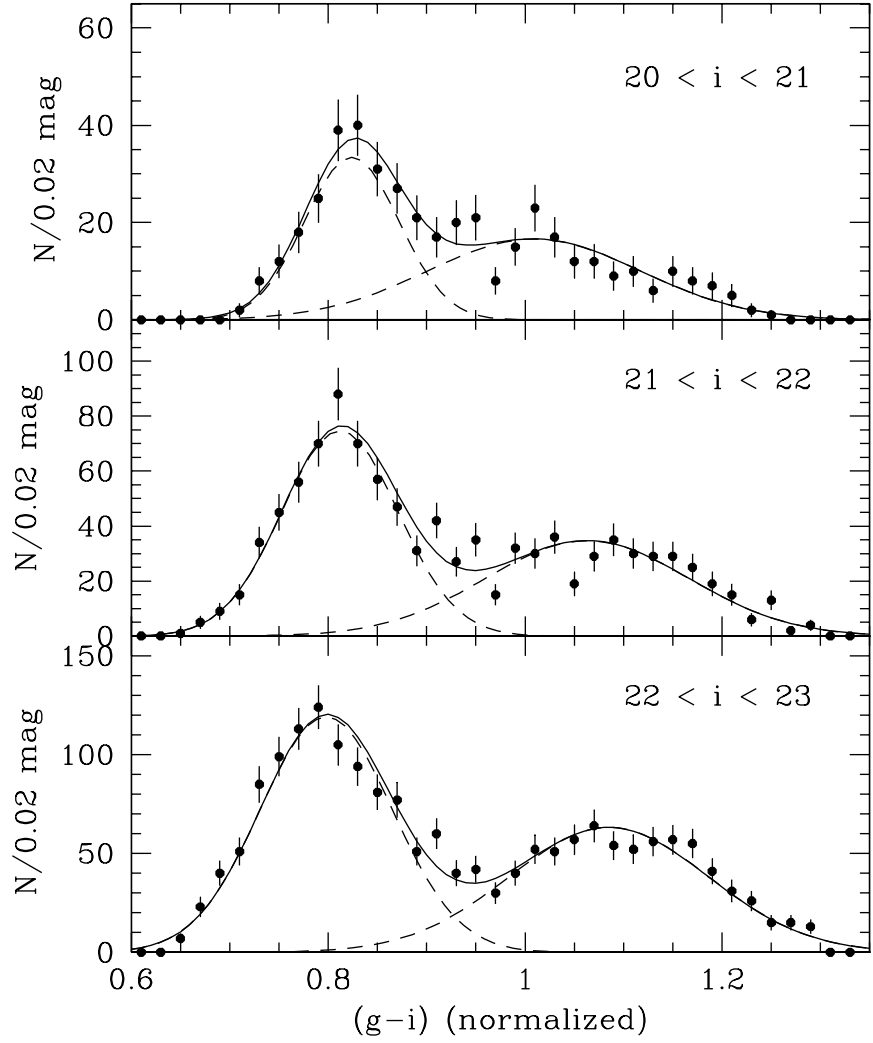}
\caption{Sample RMIX solutions for the $(g-i)$ color distributions in
three one-magnitude bins in $i$ as labelled.  The $(g-i)$ colors 
of the individual clusters are normalized
to a galactocentric distance $R = 1.0 R_{eff}$ to remove the spatial metallicity
gradients in the system (see text).   In each panel, the dashed lines show
the Gaussian curves matching the blue and red sequences, while the solid lines
show the sum of the two components.
}
\label{rmix_histo}
\end{figure}

As an illustration of the overall quality of the fundamental assumption
of bimodality, 
sample histogram fits for three representative magnitude
ranges are shown in Figure \ref{rmix_histo}.
Two modes are definitely required, but the arbitrary addition of more
than two subcomponents results in no formal improvement to the total
color distribution \citep[see also][for similar tests in several other galaxies]{lar01,har09,coc09},
showing that bimodality is both a necessary and sufficient condition 
for describing the broadband-color distribution.

\subsection{The Blue Sequence}

\begin{figure}
\plotone{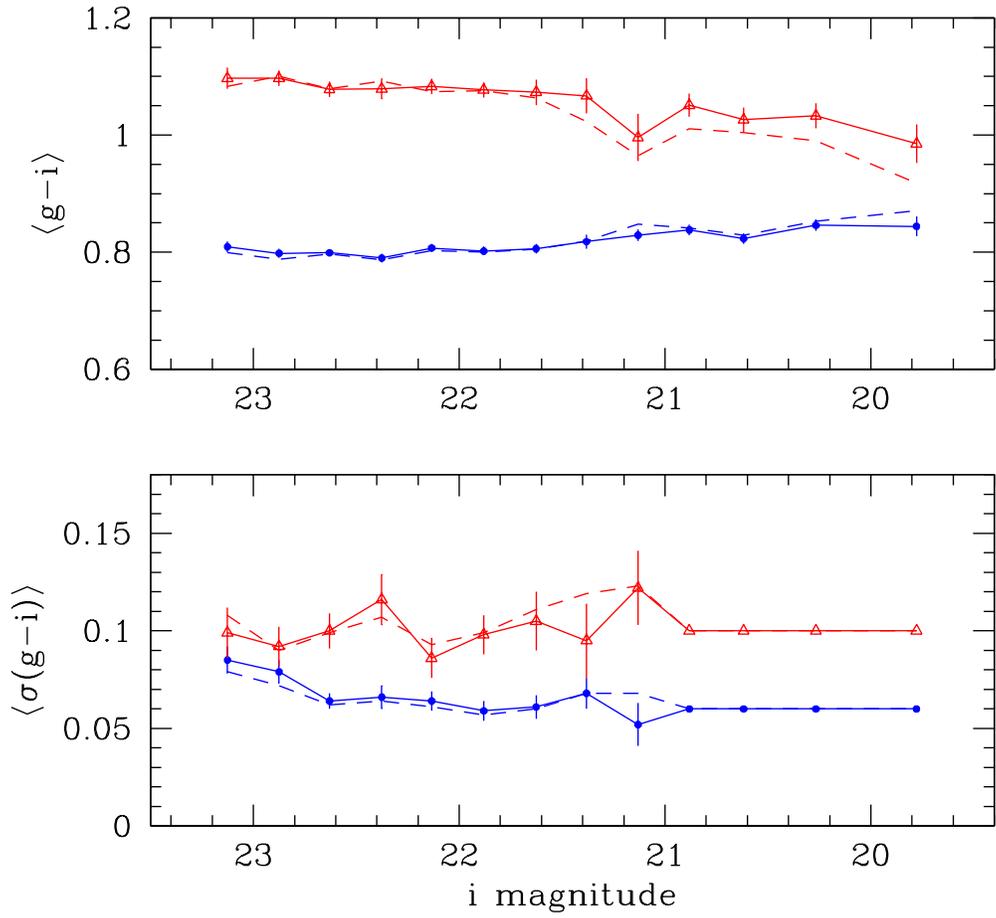}
\caption{\emph{Upper panel:} Binned mean colors in $(g-i)$ for the blue (solid
dots) and red (open triangles)
sequences, as listed in Table \ref{rmixblue}.   \emph{Lower panel:} Internal
standard deviation (color width) of the blue and red sequences as a function
of $i$ magnitude.  For $i<21$, the $\sigma-$values are constrained at their
mean levels (see text).  
For the blue sequence (lower curve in each panel), the \emph{dashed lines}
show how the fitted color and dispersion change if the red sequence is
constrained to have $\mu_2 \equiv 1.07$ at all magnitudes, as listed
in Table 3.
For the red sequence (upper curve in each panel), the dashed lines
show the changes if the blue sequence is constrained to have 
color $\mu_1 \equiv 0.80$ at all magnitudes.
}
\label{meancolors}
\end{figure}

The behavior of the key bimodal parameters (the means $\mu_1, \mu_2$
and the dispersions $\sigma_1, \sigma_2$) with luminosity is shown in
Figure \ref{meancolors}.  A nonzero MMR will appear as a systematic
trend of the mean color $\mu$ with magnitude.  By using the bimodal
fits to define the central colors of each mode, we also explicitly account for 
any overlap of the color distributions and possible asymmetric biasing
due to contamination of one mode by the other.
We concentrate first on the trends along
the blue sequence, which to date is the only one of the two claimed
to show a clear MMR, from the studies of many other galaxies.  

The \emph{blue sequence} in M87 shows up as a narrow, rather sharply
defined feature in the CMD that is easily traced over the entire
spatial region studied here.  
The narrow spread and large population on the blue sequence mean that
the numerical solution for its position and shape is
very stable.  In the subsample used here for the fits, 
it takes up 55-60\% of the total GC population, and the $p_1$ ratio
stays nearly constant at all magnitude levels 
(upper panel of Figure \ref{chifrac}).\footnote{This $p_1$ value is
only a local one for the spatial region used here.  The global one over
the entire M87 halo will be larger because the blue GCs extend
much further out into the halo.  For comparison, \citet{pen08} find
$p_1$(global) $= 0.73$ for M87 by integrating the number counts 
for both types of clusters to large radius.}
Comparison of the results in Tables
2 and 3 (and displayed in Figure \ref{meancolors}) shows that no significant
changes in $\mu_1$ or $\sigma_1$ occur when the red sequence is arbitrarily
required to fall at a constant color.  
The mean color (normalized to $R = 1.0~ R_{eff}$)
is $\langle g-i \rangle(blue) = 0.80 \pm 0.01$ over the fainter half of the
$i-$magnitude range studied here.  Its intrinsic width has a
dispersion $\sigma_1 = 0.06 \pm 0.01$ that remains highly
consistent with magnitude, 
except in the faintest range $i > 22.5$  where it is partially
broadened by photometric measurement scatter.  
The \citet{mar05} population models used above then show that
the dispersion in color corresponds
to an intrinsic rms metallicity spread $\sigma$(Fe/H) = $0.30 \pm 0.08$.
For comparison, \citet{har09} found $\sigma$(Fe/H)(blue) = 0.27 for the 
average of six other BCGs, measured from the $(B-I)$ color index.

The bimodal fits show that the blue sequence becomes
smoothly and distinctly redder with increasing luminosity, especially for 
$i \lesssim 21.5$ ($M_I \lesssim -10$).  Over this range, the
mean color behaves as $(g-i)_{norm} = (0.845 \pm 0.090) - (0.021 \pm 0.004)~(i-20)$,
a slope that is different from zero at the $5\sigma$ level.
The resultant scaling of GC heavy-element
abundance with luminosity
is then $Z \sim L^{0.25 \pm 0.08}$.  
For comparison, \citet{coc09} and \citet{har09} find an average blue-sequence scaling
of $Z \sim L^{0.3\pm0.05}$ for more than a dozen large E galaxies measured
from a variety of both ground-based and HST data and applying to
a very similar GC luminosity range, though these studies also indicate
that real differences from one galaxy to the next may exist.  

At fainter levels ($M_I \gtrsim -10$, corresponding roughly to $10^6 M_{\odot}$), 
it is not clear that any significant trend exists and that the blue sequence
could better be considered vertical.  In summary, these results for M87
are closely similar to those of \citet{har09} for six other BCG giants:
the MMR for the blue sequence in M87 exists, but it is nonlinear in form and
most strongly affects the most massive clusters.  Comparison of the
Megacam result with the recent thorough analysis of the deep $(V-I)$
data from HST/ACS \citep{pen09} also shows close similarities:
they found a clearly nonzero trend towards redder colors over the magnitude
range $M_I \lesssim -9$ and a nearly vertical sequence below that.

The data discussed here emphasize that detecting the 
MMR correlation on observational grounds
depends not just on using a metallicity-sensitive photometric index, but also 
on having a \emph{very large sample that extends to high luminosity}.
If we had restricted our analysis to just the inner zone (see, for example,
the first panel of Figure \ref{cmd_zone}), the MMR with its nonlinear form would be essentially
undetectable.  The \citet{wat09} study, which uses $(V-I)$ colors for the
clusters that fall in a single HST/ACS field, covers only the region within 
$2'$ of the center of M87.  Their conclusion that no MMR is present in fact does not disagree
with the present Megacam results for two reasons:  first, at lower GC luminosity
$M_I \gtrsim -10$ where the vast majority of the clusters fall, the MMR actually
does become more nearly vertical 
and no trend is expected.  Second, the smaller sample size
in the HST/ACS data means that only a few dozen of its clusters 
lie above the transition level $M_I \sim -10$ where the MMR starts to
become significant, making it hard to distinguish above the intrinsic
scatter in the sequence.  A more detailed comparative
analysis of the HST/ACS data is provided by \citet{pen09}.

\subsection{The Red Sequence}

The \emph{red sequence} has a typical mean color 
$(g-i)_{norm} = 1.07$ averaged over all magnitudes, but it is
less precisely definable at any level because its intrinsic
dispersion $\sigma_2 \simeq 0.10 \pm 0.01$ is nearly twice as
broad as that of the blue sequence.  This larger spread also makes it
more vulnerable to the field contamination at the upper end. 
The corresponding metallicity
spread is $\sigma$(Fe/H) = $0.50 \pm 0.13$.  From their $(C-T_1)$ photometry,
\citet{for07} measure a ratio $\sigma_1/\sigma_2 = 0.60$, in
exact agreement with ours.  For a combination of six other BCG galaxies,
\citet{har09} found $\sigma$(Fe/H)(red) = 0.45 from $(B-I)$, also
in good agreement with the M87 sequences.

In previous studies with other galaxies 
\citep[notably][where explicit tests have been made]{har06,mie06,har09,coc09,pen09}
no significant MMR's have been claimed to exist along the red sequence.
On theoretical grounds, if self-enrichment is the physical origin of
the MMR then we do not expect to find any correlation
of metallicity with mass along the red sequence \emph{except} for extremely
massive clusters (several million Solar masses and higher)
where the amount of self-enrichment can finally reach above
the already-large pre-enrichment level \citep[see][]{bai09}.
In most galaxies neither the blue nor red sequences reach that high.
However, the large and homogeneous database for M87 offers a chance
to test this notion empirically.

For the fainter range $i \gtrsim 21.5$, no significant change
shows up in the mean color $\mu_2$.  At brighter levels, the formal
solutions show a surprisingly large change towards \emph{bluer}
mean color at higher luminosity; that is, the two sequences give
the appearance of converging towards each other there.  Taken at face
value, the red sequence would then display a ``negative MMR'' towards
lower metallicity at higher mass, contrary to any existing model
interpretation.  Nominally, the slope for the unconstrained
fits in Table 2 is $\Delta(g-i) / \Delta i \simeq 0.03$,
corresponding to $Z \sim L^{-(0.3\pm0.15)}$. 

Closer examination indicates, however, that this trend is
not as robust as for the blue sequence.
Formally the slope is significant at only the $2-\sigma$ level.  
Perhaps more importantly, the combined effects of 
its higher internal spread and the significant amount of field
contamination in the CMD make the mean position of the sparsely
populated upper end of the red sequence quite uncertain.  
This is shown in Table 4 and the upper panel of Figure \ref{meancolors}.
The numerical tests described above
show that constraining the position of the \emph{blue} sequence $\mu_1$ has a
large effect on the deduced position of $\mu_2$, pulling it much
further to the blue.  By contrast, the opposite test of
constraining the red sequence
color has no important effect on the blue sequence (lower pair 
of lines in Figure \ref{meancolors}).  

\begin{figure}
\plotone{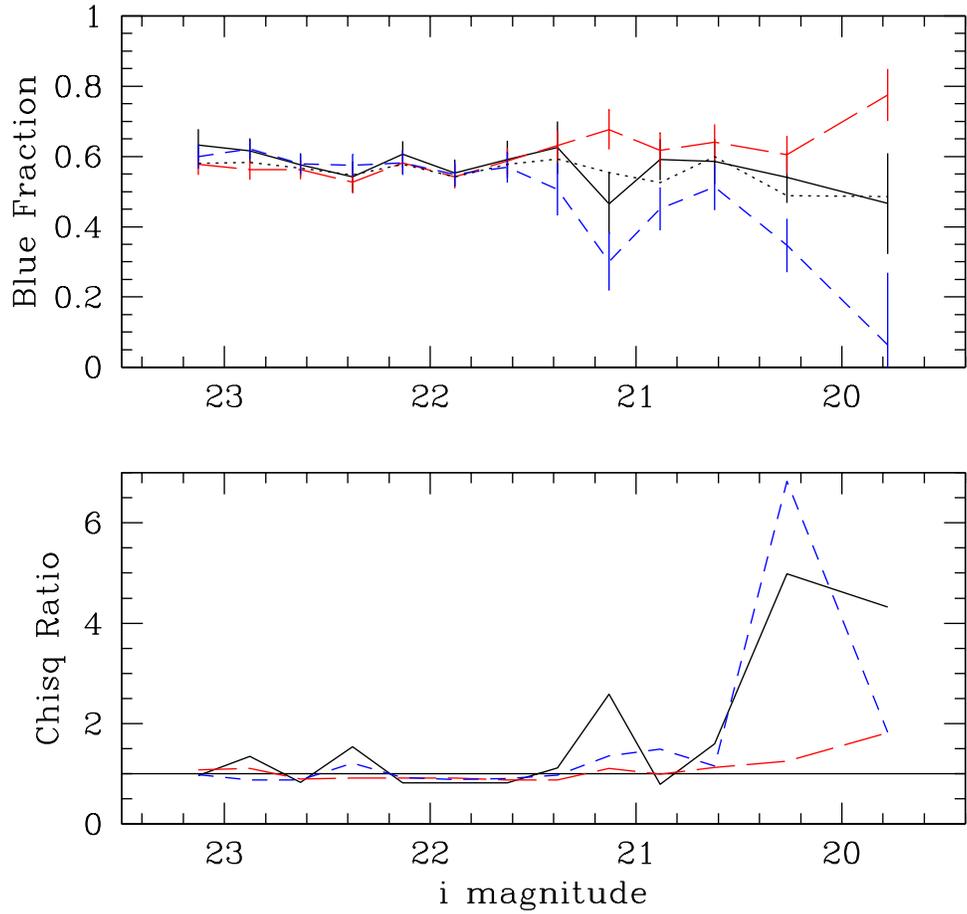}
\caption{\emph{Upper panel:} The fraction $p_1$ of the total GC population
in the blue sequence, plotted as a function of $i$ magnitude and obtained
from the bimodal RMIX fits.  The solid line shows the results from Table 2,
where both blue and red means $\mu_1, \mu_2$ are unconstrained.  The red long-dashed line
shows the results where the red sequence is constrained to lie at $\mu_2 \equiv
1.070$; the blue short-dashed line shows the results where the blue sequence is
constrained at $\mu_1 \equiv 0.80$; and the dotted line shows the results
where both sequences are constrained.
\emph{Lower panel:}  The quality-of-fit ratio 
$\chi_{\nu}^2$(constrained)/$\chi_{\nu}^2$(unconstrained) for three different
versions of the bimodal-sequence fits.  The solid line shows the ratio for
the case where both sequences are constrained at $\mu_1 = 0.80, \mu_2=1.07$;
the red long-dashed line shows the case for $\mu_2 = 1.07$; and the blue 
short-dashed line shows
the case for $\mu_1 = 0.80$.  A ratio larger than 1.00 means that the 
solution provides a worse fit than the unconstrained case.
}
\label{chifrac}
\end{figure}

Another result of these tests is shown in the lower panel of Figure \ref{chifrac}.  
Here, the ratio $\chi^2$(constrained)/$\chi^2$(unconstrained) is plotted
for the different sets of bimodal fits, to demonstrate what happens to
the goodness-of-fit of each solution relative to the ``baseline'' model
fit of Table 2.  These $\chi^2$ ratios are listed in the last columns
of Tables 3 and 4.  If we arbitrarily fix the red sequence (Table 3, and red
line in Figure \ref{chifrac})), then the quality of fit remains virtually
as good as the baseline model except for the very brightest bin.
But for the opposite case where we fix the blue sequence (Table 4, and
blue line in Figure \ref{chifrac}), a much more dramatic change occurs:
at the bright end, the quality of fit degrades strongly.  
The same is true for the final set of fits where both sequences are
arbitrarily constrained to their mean values.  
The case where the blue sequence is fixed also leads to a steep and
unrealistic decrease in the $p_1$ ratio at the bright end (blue line in the
upper panel of Figure \ref{chifrac}).

In other words, over the range $i \gtrsim 21$ ($M_I \gtrsim -10.5$),
the assumption that both sequences are strictly vertical (no MMR) provides
an entirely reasonable fit to the data.  Brighter than this,
the null hypothesis of no MMR becomes increasingly invalid, and
closer attention must be paid to the curvature in one or both
of the sequences.

In summary, the present data point clearly to the existence of
a blue-sequence MMR.  Correctly representing it 
is crucial for the quality of fit of the entire color distribution.
By contrast, the data indicate -- though much more 
hesitantly -- that a negative MMR may exist along the red sequence,
though we cannot reject the null hypothesis (no MMR) as yet:
the numerical tests described above show that ignoring the red MMR
has very little effect on the quality of the bimodal fits.

The recent study of \citet{pen09} uses a sample of $\sim 2000$ M87
clusters measured in the less sensitive $(V-I)$ color, but has the
advantage that it is almost completely free of field contamination.
They find that the red sequence has no significant slope at all
luminosity levels.

\section{Discussion and Interpretation}

The results of this study fall into line
with the most complete recent collections of data for other
large galaxies \citep{mie06,har09,coc09}.  Above a transition
point of about $10^6 M_{\odot}$, the low-metallicity
clusters become progressively more enriched at higher luminosity, 
scaling as $Z \sim L^{0.25}$.  This nonlinear MMR can be reasonably
well understood through a basic self-enrichment model \citep{str08,bai09}.

At the same time, a MMR of similar amplitude and reverse direction --
but of much lower significance -- is found along the red sequence,
$Z \sim L^{-(0.3 \pm 0.15)}$.  Interestingly,   
\citet{coc09} find a mean scaling $Z$(red sequence) $ \sim L^{-(0.1 \pm 0.1)}$
for the average of 15 giant E galaxies, a result that is only marginally different
from zero but at least points in the same direction as the M87 data.
Thus it may be worth considering how the red sequence
could have embedded in it some form of substructure appearing as a MMR of its own
in a way that does \emph{not} require self-enrichment.

To discuss the possibilities, we first ask what the essential differences are between the
two modes.
The blue, metal-poor clusters make up a narrow, well defined sequence that is
replicated in very much the same way over and over in galaxies of all types, differing
primarily just in their total numbers within different galaxies.  Their mean metallicity 
$\langle$Fe/H$\rangle \simeq -1.4$ is
only weakly correlated with galaxy mass and shows very much the
same internal dispersion $\sigma$(Fe/H) $\simeq 0.3$ from one system to another 
\citep[e.g.][]{bur01,str04,bro06,pen06}.  In addition, the blue
GCs display a consistently
old $\sim 13-$Gy mean age with little internal age spread
\citep[e.g.][]{puz05,bro06,bea08,mar09}.  Many lines of argument advanced
in recent years have built up the view that the metal-poor GCs are likely to have
formed in the first brief round of star formation within the ``pregalactic
dwarfs'' that began their star formation
just at the beginning of the major stage of hierarchical merging,
and additional metal-poor GCs could have
accreted later as part of their host dwarf satellites
\citep[see especially][among many others]{sea78,har94,for97,cot98,bur01,bea02,kra05,bro06,bek08}.
These small host dwarfs with their shallow potential wells and initially metal-poor gas
could support only one major round of star formation and thus left behind metal-poor
stars plus much unconverted gas.

By contrast, the red, metal-richer sequence is a more 
heterogeneous population that in some ways is harder to model in detail.
Their mean metallicity $\langle$Fe/H$\rangle \simeq -0.4$ correlates
weakly with host galaxy luminosity \citep{for97,str04,bro06}.  Both their
metallicity distribution and spatial distribution within the halos of
their galaxies as a whole are similar to the 
\emph{field-star} population that has a broad, metal-rich MDF and makes
up the bulk of a typical giant elliptical \citep{gei96,har02,rho04,for07}. That is,
the metal-rich GCs can reasonably be thought of as forming
during the major 
stage of hierarchical merging that built most of the galaxy.  
Hierarchical models \citep[e.g.][]{bea02,bea03} show that multiple
``in situ'' starbursts, major mergers, and accretions of gas-rich satellites
would all have contributed to this stage.  If this basic picture is correct,
then the final population of
metal-rich GCs would be expected to have a wide range of ages and 
metallicities, and more so in bigger galaxies that experienced more starbursts
and merging events.
Furthermore, the red-GC populations should also differ in detail
from one galaxy to another depending on their
individual merger and starburst histories.

On the observational side, the evidence points increasingly to the same view that
the metal-rich GC ``population'' is actually a composite of objects with
various ages, with its subcomponent MDFs heavily overlapped and 
thus of nearly similar mean metallicities 
\citep[e.g.][]{puz05,kun05,bro05,dir05,lar05,bro06,cen07,gou07,bea08,may09}.  
Consistent with this view, both the relative numbers and luminosity functions of the red sequence
differ noticeably from one galaxy to another, so that
the high-luminosity end of the red
GCLF does not extend to the same maximum even in galaxies with similar
total GC populations.
For example, in NGC 1399 the blue and red sequences have
very similar ``upper ends'' while in M87 the blue sequence extends much higher
\citep{for07}; see also the difference between the Antlia giants NGC 3258 and 3268 
\citep{bas08}.  By contrast with most other systems, 
in the Hydra and Coma supergiants NGC 3311 and NGC 4874
the red sequence is the one that extends to higher luminosity, up to cluster
masses $\simeq 10^7 M_{\odot}$ and perhaps higher \citep{weh08,har09b}.  
Observations of young, massive star clusters in present-day starburst galaxies
\citep{lar00,lar09} also show that the cluster formation efficiency and the
initial cluster mass function can vary with the total star formation rate.

These various pieces of evidence point to a way to understand, 
at least in principle, how we might obtain
a red GC sequence with an observable mass/metallicity relation
that could be \emph{either} positive or negative.
In the simplest such case, suppose that the metal-rich GCs comprise
two main subpopulations 
that can be thought of as the products of the two biggest mergers or starbursts
that assembled the main body of the galaxy.
In addition, suppose that the earlier and slightly \emph{less} metal-rich
of these two starbursts generated its GCs with a broader mass function that extended to
higher luminosity than did the later and metal-richer burst.
What we would see today in the GCS as a whole would then be a 
red sequence with a broad MDF and a ``negative MMR'' at the top end. 
This population-based approach has nothing to do with any
self-enrichment, is completely consistent with the
evidence summarized above, and does not invoke any unusual physical effect.
Simulated GCSs along these lines are now being constructed to test
these ideas further and will be discussed in a later paper.

\section{Summary}

Wide-field photometry with CFHT/Megacam in $gri$ has been used to
study the properties of the M87 globular cluster system.
Approximately 6200 GCs have been measured to a limit
$i \simeq 23.0$ (equivalent roughly to $M_I \sim -8.5$, a little
less than half the total population).  
The primary results of this study are these:
\begin{enumerate}
\item{} The metal-poor (blue) and metal-rich (red) cluster
sequences are clearly delineated especially in the $(g-i)$ color
index. While the red sequence does not extend detectably outward
to projected galactocentric distances beyond $\sim$50 kpc against
the field contamination, the blue sequence can be seen to 100 kpc
and beyond.
\item{} Both the red and blue subpopulations have shallow but
significant radial metallicity gradients, corresponding to 
heavy-element abundance scaling with distance of
$Z \sim R^{-0.12}$ (blue) and $Z \sim R^{-0.17}$ (red).
\item{} The blue sequence shows a nonlinear mass/metallicity
relation:  GC metallicity increases with luminosity.
This relation is nearly vertical (that is, mean color
is uncorrelated with metallicity) for $M_I \gtrsim -10$,
while at brighter levels we find a trend roughly
reproduced by $Z \sim L^{0.25}$.
This result agrees well with the findings in numerous
other giant E galaxies, as well as the best current
data on M87 itself. Finding the trend depends critically on being able
to sample the GC population to its highest luminosity levels.
The blue-sequence MMR can be physically understood 
as the result of GC self-enrichment during formation.
\item{} The red sequence is almost twice as broad as the
blue sequence in its internal scatter of colors.   The 
formal numerical solutions show that it has a ``negative MMR'' at
the top end that is at least as steep as the blue-sequence 
MMR.  However, the same numerical solutions show that this
trend is not strongly significant, and simply assuming 
a constant mean color for the red sequence does not noticeably
decrease the quality of the bimodal fits.
\item{} In giant galaxies like M87, the red-GC sequence is
likely to be a composite population, unlike the more
internally homogeneous blue sequence.  Every major starburst
or gas-rich merger that contributed to the formation of the
main body of the galaxy should have contributed to its
metal-rich GC population.  If the individual major starbursts
generated GCs following slightly different initial mass
functions, then features such as the negative-MMR can be
built in to the total red sequence that we see today.
\end{enumerate}

\acknowledgments
This work was supported by the Natural Sciences and Engineering
Research Council of Canada through research grants to WEH,
and by the Killam Foundation of the Canada Council through a 
research fellowship.
It is a pleasure to acknowledge the Canadian Astronomy
Data Center (CADC) and the high quality of their archive products
that facilitated this study.


\clearpage


\begin{deluxetable}{cccc}
\tabletypesize{\footnotesize}
\tablecaption{CFHT/Megacam Images\label{images}}
\tablewidth{0pt}
\tablehead{
\colhead{Filter} & \colhead{$N$(exp)} & \colhead{Total t (sec)} & \colhead{FWHM} \\
}
\startdata
$g'$ & 11 & 3520 & $0.74''$ \\
$r'$ & 10 & 3000 & $0.80''$ \\
$i'$ & 11 & 3300 & $0.82''$ \\
\\
\enddata

\end{deluxetable}

\begin{deluxetable}{cccc}
\tabletypesize{\footnotesize}
\tablecaption{Radial Distributions\label{radialdat}}
\tablewidth{0pt}
\tablehead{
\colhead{$R$ Range} & \colhead{$R$} & \colhead{$n_{cl}$} & 
\colhead{$\sigma_{cl}$ (arcmin$^{-2}$) } \\
}
\startdata
Blue GCs \\
 $  1.00'-1.30'$  & $1.14'$ & 68.4 & $31.57 \pm  3.83$  \\
 $  1.30-1.69$  & 1.48   &   84.1 & $22.94 \pm  2.52$ \\
 $  1.69-2.20$  & 1.93   &  128.4 & $20.75 \pm  1.84$ \\
 $  2.20-2.86$  & 2.50   &  135.3 & $12.93 \pm  1.13$  \\
 $  2.86-3.71$  & 3.26   &  206.4 & $11.67 \pm  0.83$  \\
 $  3.71-4.83$  & 4.23   &  256.2 & $ 8.57 \pm  0.55$  \\ 
 $  4.83-6.27$  & 5.50   &  293.9 & $ 5.82 \pm  0.36$  \\ 
 $  6.27-8.16$  & 7.15   &  388.8 & $ 4.56 \pm  0.26$  \\ 
 $  8.16-10.60$ &  9.30  &   428.5& $ 2.97 \pm  0.18$  \\ 
 $ 10.60-13.79$ & 12.09  &   423.6& $ 1.74 \pm  0.14$  \\ 
 $ 13.79-17.92$ & 15.72  &   426.8& $ 1.04 \pm  0.12$  \\ 
 $ 17.92-23.30$ & 20.43  &   472.9& $ 0.68 \pm  0.11$ \\ 
 $ 23.30-30.29$ & 26.56  &   382.1& $ 0.33 \pm  0.10$ \\ 
\\
Red GCs \\
 $  1.00'-1.30'$  & $1.14'$ & 81.2 & $37.47 \pm  4.18$  \\
 $  1.30-1.69$  & 1.48   &  102.7 & $28.03 \pm  2.79$ \\
 $  1.69-2.20$  & 1.93   &  135.8 & $21.93 \pm  1.90$ \\
 $  2.20-2.86$  & 2.50   &  143.2 & $13.69 \pm  1.16$  \\
 $  2.86-3.71$  & 3.26   &  167.6 & $ 9.48 \pm  0.75$  \\
 $  3.71-4.83$  & 4.23   &  170.2 & $ 5.70 \pm  0.46$  \\ 
 $  4.83-6.27$  & 5.50   &  186.8 & $ 3.70 \pm  0.30$  \\ 
 $  6.27-8.16$  & 7.15   &  189.3 & $ 2.22 \pm  0.20$  \\ 
 $  8.16-10.60$ &  9.30  &   187.1& $ 1.30 \pm  0.15$  \\ 
 $ 10.60-13.79$ & 12.09  &   166.2& $ 0.68 \pm  0.12$  \\ 
 $ 13.79-17.92$ & 15.72  &   145.7& $ 0.35 \pm  0.11$  \\ 
 $ 17.92-23.30$ & 20.43  &    94.4& $ 0.14 \pm  0.10$ \\ 
\\
\enddata  

\end{deluxetable}

\clearpage
\begin{deluxetable}{ccccccc}
\tabletypesize{\footnotesize}
\tablecaption{RMIX Bimodal Fits to the $(g-i)$ Colors\label{rmixblue}}
\tablewidth{0pt}
\tablehead{
\colhead{$i$ range} & \colhead{$N$(bin)} &
\colhead{$\mu_1~(\pm)$} & \colhead{$\sigma_1~(\pm)$} &  \colhead{$\mu_2~(\pm)$} & 
\colhead{$\sigma_2~(\pm)$} & \colhead{$p_1 ~(\pm)$} \\
}
\startdata
$(23.00, 23.25)$ &455 & $0.809 (0.009) $ & $0.085 (0.007) $ & $1.097 (0.018)$ & $0.099 (0.013) $ & $0.633 (0.035)$ \\
$(22.75, 23.00)$ &497 & $0.798 (0.008) $ & $0.079 (0.006) $ & $1.097 (0.013)$ & $0.092 (0.010) $ & $0.616 (0.035)$ \\
$(22.50, 22.75)$ &486 & $0.799 (0.006) $ & $0.064 (0.004) $ & $1.078 (0.013)$ & $0.100 (0.009) $ & $0.576 (0.035)$ \\
$(22.25, 22.50)$ &397 & $0.790 (0.007) $ & $0.066 (0.006) $ & $1.079 (0.017)$ & $0.116 (0.013) $ & $0.542 (0.035)$ \\
$(22.00, 22.25)$ & 330 & $0.807 (0.007) $ & $0.064 (0.005) $ & $1.083 (0.013)$ &$0.086 (0.010) $ & $0.606 (0.035)$ \\
$(21.75, 22.00)$ & 311 & $0.802 (0.006) $ & $0.059 (0.005) $ & $1.077 (0.013)$ &$0.098 (0.010) $ & $0.553 (0.035)$ \\
$(21.50, 21.75)$ & 257 & $0.806 (0.008) $ & $0.061 (0.006) $ & $1.073 (0.022)$ & $0.105 (0.015) $ & $0.592 (0.035)$ \\
$(21.25, 21.50)$ & 191 & $0.818 (0.012) $ & $0.068 (0.008) $ & $1.067 (0.030)$ & $0.095 (0.019) $ & $0.625 (0.035)$ \\
$(21.00, 21.25)$ & 152 & $0.829 (0.010) $ & $0.052 (0.011) $ & $0.996 (0.040)$ & $0.122 (0.019) $ & $0.415 (0.035)$ \\
$(20.75, 21.00)$ & 138 & $0.838 (0.009) $ & $0.060 (0.000) $ & $1.051 (0.020)$ & $0.100 (0.000) $ & $0.591 (0.058)$ \\
$(20.50, 20.75)$ & 115 & $0.823 (0.009) $ & $0.060 (0.000) $ & $1.026 (0.021)$ & $0.100 (0.000) $ & $0.586 (0.064)$ \\
$(20.00, 20.50)$ & 122 & $0.846 (0.010) $ & $0.060 (0.000) $ & $1.033 (0.021)$ & $0.100 (0.000) $ & $0.540 (0.072)$ \\
$(19.50, 20.00)$ &  57 & $0.844 (0.017) $ & $0.060 (0.000) $ & $0.985 (0.033)$ & $0.100 (0.000) $ & $0.466 (0.143)$ \\
\\
\enddata

\end{deluxetable}

\begin{deluxetable}{ccccccc}
\tabletypesize{\footnotesize}
\tablecaption{Bimodal Fits (Red Sequence Constrained)\label{rmixv2}}
\tablewidth{0pt}
\tablehead{
\colhead{$i$ range} & \colhead{$N$(bin)} &
\colhead{$\mu_1~(\pm)$} & \colhead{$\sigma_1~(\pm)$} &  
\colhead{$\sigma_2~(\pm)$} & \colhead{$p_1 ~(\pm)$} & \colhead{$\chi_{\nu}^2(con)/\chi_{\nu}^2(uncon)$} \\
}
\startdata
$(23.00, 23.25)$ &455 & $0.799 (0.009) $ & $0.079 (0.005) $ &  $0.114 (0.009) $ & $0.578 (0.030)$ & 1.08 \\
$(22.75, 23.00)$ &497 & $0.788 (0.006) $ & $0.072 (0.005) $ &  $0.108 (0.007) $ & $0.562 (0.028)$ & 1.11 \\
$(22.50, 22.75)$ &486 & $0.797 (0.005) $ & $0.062 (0.004) $ &  $0.104 (0.007) $ & $0.562 (0.027)$ & 0.90 \\
$(22.25, 22.50)$ &397 & $0.787 (0.006) $ & $0.064 (0.005) $ &  $0.121 (0.009) $ & $0.527 (0.031)$ & 0.92 \\
$(22.00, 22.25)$ & 330 & $0.803 (0.006) $ & $0.061 (0.004) $ & $0.093 (0.008) $ & $0.582 (0.032)$ & 0.91 \\
$(21.75, 22.00)$ & 311 & $0.800 (0.005) $ & $0.057 (0.004) $ & $0.102 (0.008) $ & $0.542 (0.033)$ & 0.92 \\
$(21.50, 21.75)$ & 257 & $0.805 (0.006) $ & $0.060 (0.005) $ & $0.106 (0.010) $ & $0.587 (0.038)$ & 0.88 \\
$(21.25, 21.50)$ & 191 & $0.819 (0.008) $ & $0.068 (0.006) $ & $0.093 (0.011) $ & $0.631 (0.043)$ & 0.88 \\
$(21.00, 21.25)$ & 152 & $0.848 (0.010) $ & $0.068 (0.008) $ & $0.094 (0.013) $ & $0.676 (0.056)$ & 1.11 \\
$(20.75, 21.00)$ & 138 & $0.841 (0.008) $ & $0.060 (0.000) $ & $0.100 (0.000) $ & $0.618 (0.048)$ & 0.99 \\
$(20.50, 20.75)$ & 115 & $0.829 (0.008) $ & $0.060 (0.000) $ & $0.100 (0.000) $ & $0.640 (0.051)$ & 1.13 \\
$(20.00, 20.50)$ & 122 & $0.853 (0.009) $ & $0.060 (0.000) $ & $0.100 (0.000) $ & $0.605 (0.054)$ & 1.25 \\
$(19.50, 20.00)$ &  57 & $0.871 (0.011) $ & $0.060 (0.000) $ & $0.100 (0.000) $ & $0.775 (0.074)$ & 1.82 \\
\\
\enddata

\end{deluxetable}

\begin{deluxetable}{ccccccc}
\tabletypesize{\footnotesize}
\tablecaption{Bimodal Fits (Blue Sequence Constrained)\label{rmixv3}}
\tablewidth{0pt}
\tablehead{
\colhead{$i$ range} & \colhead{$N$(bin)} & \colhead{$\sigma_1~(\pm)$} &
\colhead{$\mu_2~(\pm)$} & \colhead{$\sigma_2~(\pm)$} &  
\colhead{$p_1 ~(\pm)$} & \colhead{$\chi_{\nu}^2(con)/\chi_{\nu}^2(uncon)$} \\
}
\startdata
$(23.00, 23.25)$ &455 & $0.081 (0.005) $ & $1.083 (0.014) $ &  $0.108 (0.011) $ & $0.600 (0.034)$ & 0.98 \\
$(22.75, 23.00)$ &497 & $0.080 (0.005) $ & $1.100 (0.010) $ &  $0.090 (0.008) $ & $0.621 (0.027)$ & 0.88 \\
$(22.50, 22.75)$ &486 & $0.064 (0.004) $ & $1.079 (0.010) $ &  $0.099 (0.008) $ & $0.579 (0.028)$ & 0.88 \\
$(22.25, 22.50)$ &397 & $0.071 (0.005) $ & $1.092 (0.012) $ &  $0.107 (0.009) $ & $0.575 (0.031)$ & 1.22 \\
$(22.00, 22.25)$ & 330 & $0.061 (0.005) $ & $1.074 (0.012) $ & $0.093 (0.009) $ & $0.582 (0.035)$ & 0.92 \\
$(21.75, 22.00)$ & 311 & $0.058 (0.004) $ & $1.076 (0.012) $ & $0.099 (0.009) $ & $0.549 (0.034)$ & 0.88 \\
$(21.50, 21.75)$ & 257 & $0.058 (0.005) $ & $1.063 (0.018) $ & $0.111 (0.013) $ & $0.569 (0.044)$ & 0.91 \\
$(21.25, 21.50)$ & 191 & $0.059 (0.007) $ & $1.023 (0.027) $ & $0.119 (0.017) $ & $0.506 (0.074)$ & 0.98 \\
$(21.00, 21.25)$ & 152 & $0.042 (0.009) $ & $0.965 (0.019) $ & $0.123 (0.010) $ & $0.301 (0.083)$ & 1.36 \\
$(20.75, 21.00)$ & 138 & $0.060 (0.000) $ & $1.011 (0.017) $ & $0.100 (0.000) $ & $0.451 (0.061)$ & 1.49 \\
$(20.50, 20.75)$ & 115 & $0.060 (0.000) $ & $1.004 (0.020) $ & $0.100 (0.000) $ & $0.513 (0.066)$ & 1.16 \\
$(20.00, 20.50)$ & 122 & $0.060 (0.000) $ & $0.990 (0.018) $ & $0.100 (0.000) $ & $0.347 (0.076)$ & 6.83 \\
$(19.50, 20.00)$ &  57 & $0.060 (0.000) $ & $0.918 (0.028) $ & $0.100 (0.000) $ & $0.063 (0.206)$ & 1.83 \\
\\
\enddata

\end{deluxetable}

\clearpage

\clearpage

\clearpage

\clearpage

\clearpage

\clearpage

\clearpage

\clearpage

\clearpage

\begin{figure}
\plotone{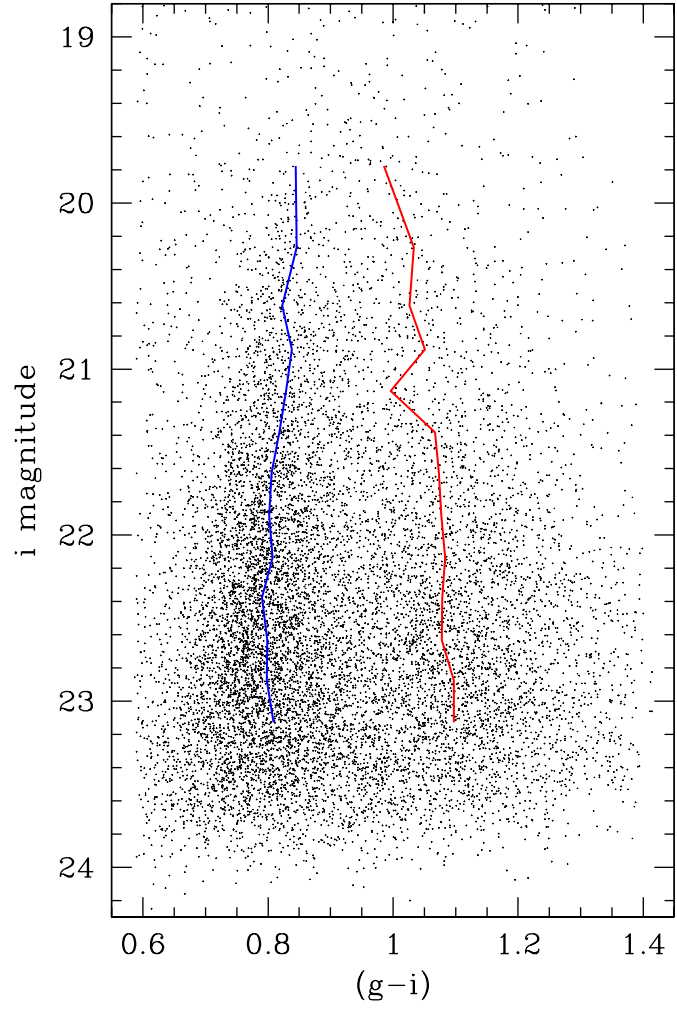}
\caption{Color-magnitude diagram in $(i,g-i)$ for the data used in
the bimodal sequence fits (see text).  The \emph{solid lines} connect
the mean points from Table \ref{rmixblue} from the unconstrained bimodal fits.
}
\label{cmdfits}
\end{figure}
\clearpage

\end{document}